\documentclass[aps,prd,twocolumn,nofootinbib,preprintnumbers,superscriptaddress]{revtex4-1}
\usepackage{slashed}
\usepackage{amsmath}
\usepackage{amssymb}
\usepackage{graphicx} 
\usepackage{hyperref}
\usepackage{tikz}
\usetikzlibrary{arrows}
\usepackage{color}
\usepackage{bm}
\usepackage{bbold}
\usepackage{overpic}
\usepackage{verbatim}

\newcommand{\be}{\begin{eqnarray}}
\newcommand{\ee}{\end{eqnarray}}

\newcommand{\bn}{\begin{enumerate}}
\newcommand{\en}{\end{enumerate}}

\def\CO{{\cal O}}

\def\t{\tau}

\def\vec#1{\bm{#1}}

\def\sl{$SL(2,\mathbf{R})$\ }

\newcommand{\qed}{\nobreak \ifvmode \relax \else
      \ifdim\lastskip<1.5em \hskip-\lastskip
      \hskip1.5em plus0em minus0.5em \fi \nobreak
      \vrule height0.75em width0.5em depth0.25em\fi}

\allowdisplaybreaks[1]

\begin{document}
\title{Unitarity and Universality in non relativistic Conformal Field theory}

\author{Sridip Pal}
\email{srpal@ucsd.edu}
\affiliation{Department of Physics, 
University of California, San Diego\\
La Jolla, CA 92093, USA}

\def\t{\tau}

\begin{abstract}
We relate the notion of unitarity of a $(0+1)$-D conformally ($SL(2,\mathbf{R})$) invariant field theory with that of a non-relativistic conformal (Schr\"odinger) field theory using the fact that \sl is a subgroup of non-relativistic conformal (Schr\"odinger) group. Exploiting \sl unitarity, we derive the unitarity bounds and null conditions for a Schr\"odinger field theory (for the neutral as well as the charged sector). In non integer dimensions the theory is shown to be non-unitary. The use of \sl subgroup opens up the possibility of borrowing results from $(0+1)$-D \sl invariant field theory to explore Schr\"odinger field theory, in particular, the neutral sector, which has otherwise been unexplored. 
\end{abstract}


\maketitle

\section{Introduction}
The conformal field theory \cite{francesco2012conformal} has a rich literature with wide application in describing physics at relativistic fixed points. Much of its armory stem from the early papers on the representation theory of $SL(2,\mathbf{R})$, a subgroup of the conformal group \cite{harish1952plancherel, pukanszky1964plancherel, luscher1976operator}. The unitarity bound along with the null condition is one of the many consequences of the representation theory of \sl algebra \cite{mack1977all}. The conformal bootstrap program also relies on knowing conformal ($SL(2,\mathbf{R})$) blocks \cite{ferrara1973tensor,polyakov1974nonhamiltonian,mack1977duality,dolan2001conformal,Rattazzi:2008pe}. The task we take up here is to use this arsenal of \sl algebra to hammer a class of non-relativistic conformal theories (NRCFT), which are \sl invariant. 

The non-relativistic conformal invariance emerges at fixed points without Lorentz invariance, in particular, in a scenario, where the symmetry involves scaling  time and space in a separate way. If the theory permits, one can have Galilean boost invariance and invariance under special conformal transformations as well. A prime example of such kind is the theory described by Schr\"odinger equation, where the maximal kinetic invariance group is the Schr\"odinger group\cite{Niederer:1972zz}. Fermions at the unitarity limit (when the S-wave scattering length $a\to\infty$) are also described by Schr\"odinger field theory\cite{Mehen:1999nd,Nishida:2010tm,Nishida:2007pj,Regal:2004zza,Zwierlein:2004zz}. Examples of approximate non-relativistic conformal field theories include systems involving ${}^{85}Rb$ \cite{Roberts:1998zz},${}^{133}Cs$\cite{Chin:2001uan},${}^{39}K$ \cite{loftus2002resonant}, deuterons \cite{Kaplan:1998tg,Kaplan:1998we} and spin chain models\cite{Chen:2017tij}.

Much like its relativistic cousin, progress has been made regarding the form of correlators and convergence of operator product expansion (OPE) in such theories for a sector with non-zero charge using the symmetry algebra only \cite{Hagen:1972pd,Niederer:1972zz,Mehen:1999nd,Henkel:1993sg,Henkel:2003pu,Nishida:2007pj,Balasubramanian:2008dm,Goldberger:2014hca, Golkar:2014mwa} (which is Schr\"odinger algebra). The state-operator correspondence invoking the harmonic potential is available for the charged sector. Nonetheless, the neutral sector has remained elusive since the representation theory along with the concept of primary and descendant breaks down for the neutral sector\cite{Golkar:2014mwa}. Thus, there is no state-operator correspondence available for the neutral sector, neither there is a proof of OPE convergence if the four point correlator involves neutral operator(s). On the other hand, physically relevant operators like Hamiltonian, number current, stress-energy tensor are neutral. This motivates us in first place to use \sl to explore the neutral sector as one can organize the operator content according to \sl representation, which is applicable to both the neutral as well as the charged sector. To our favor, it so turns out that \sl provides strong constraints on properties of Schr\"odinger field theories even for the charged sector on top of solving all the puzzles mentioned before in context of the neutral sector.

The purpose of this work is multifold. The most important point that we make is that \sl establishes a powerful and novel link between $(0+1)$-D conformal field theory (CFT) and NRCFTs. Thus results proven for $(0+1)$-D CFTs immediately apply to NRCFTs and vice versa. In fact, using $SL(2,\mathbf{R})$, we come up with state-operator map, subsequently, derive the unitarity bound, the null condition for the neutral sector for the first time. Secondly, we reformulate the notion of unitarity in the charged sector and re-derive the unitarity bound without invoking the standard map to harmonic oscillator. This, in turn helps us to identify the non-unitary sector in fractional dimensions, which has otherwise not been known previously.  Moreover, we explore the universal features of Schr\"odinger field theories including the convergence of the operator product expansion (OPE) in the neutral sector. Convergence of OPE in all the sectors also opens up the possibility of bootstrapping these theories. We deduce the universal behavior of three point coefficient and establish for the first time that even in NRCFT, there exist infinite number of \sl primaries. In short, we explicitly unveil a complete equivalence between correlators of NRCFTs on $(\t,\vec{0})$ slice and $(0+1)$-D CFTs via the notion of \sl primaries and descendants. Last but not the least, the use of \sl primaries/descendants proves to be quintessential in operator counting of heavy particle effective field theory, where neutral scalar operators appear in the Lagrangian. Only with the aid of $SL(2,\mathbf{R})$, it is possible to organize the operator basis of heavy particle effective field theory in Schr\"odinger representation \cite{Kobach:2018nmt}. 

The paper is organized as follows. In sec.~\ref{sec:unitary}, we derive the unitary bounds and null conditions for both the charged sector as well as the neutral sector of Schr\"odinger algebra. Non-unitarity in non integer dimensions has been explored in sec.~\ref{nonuni}. The sec.~\ref{universal} deals with the universality, in particular, the OPE convergence, the asymptotic behavior of three point coefficients in Schr\"odinger invariant field theory. We conclude with an elaborate discussion pointing out potential avenues of future research. To aid the main flow of the paper, the details of \sl invariant theory and representation of Schr\"odinger algebra have been relegated to appendix~\ref{app1} and \ref{app2} respectively. The role of time reversal and parity is elucidated in appendix~\ref{app3}. The appendix~\ref{app4} expounds on defining the Euclidean Schr\"odinger field theory, which comes out as a byproduct of organizing the operator content of Schr\"odinger field theory according to \sl algebra. 


\section{Unitarity bounds \& Null conditions}\label{sec:unitary}
The Schr\"odinger group acts on space-time as follows \cite{Hagen:1972pd,Niederer:1972zz,Mehen:1999nd,Henkel:2003pu}: 
\begin{align}
t\mapsto \frac{at+b}{ct+d}\,,\quad \vec{r}\mapsto \frac{\mathbb{R}\vec{r}+\vec{v}t+f}{ct+d}\,.
\end{align}
where $ad-bc=1$, $\mathbb{R}$ is a d dimensional rotation matrix, $\vec{v}$ denotes the Galilean boost and $f$ is a spatial translation. For the sector with non-zero charge, the representation is built by translating all the operators to the origin and considering the little group generated by dilatation operator $D$, Galilean boost generator $K_i$, and special conformal transformation generator $C$. The highest weight states ($\phi_\alpha$) are annihilated by $C$ and $K_i$ i.e. 
\begin{equation}\label{schp}
[C,\phi_\alpha(0,\vec{0})]=0\,,\quad [K_i,\phi_\alpha(0,\vec{0})]=0\,.
\end{equation}
These are called primary operators. The commutators with $D$ and particle number symmetry generator $\hat{N}$ dictate the charge and the dimension of these operators $\phi_\alpha$ i.e. $[D,\phi_\alpha(0,\vec{0})]=\imath\Delta_\alpha\phi(0,\vec{0})$ and $[\hat{N},\phi_\alpha(0,\vec{0})]=N_\alpha\phi_\alpha(0,\vec{0})$.
The time and space translation generators $H$ and $P_i$ create descendant operators by acting upon primary operators and raising the dimension by $2$ and $1$ respectively. The concept of primaries and descendants breaks down within the neutral sector. Since $K_i$ and $P_j$ commute in this sector, $P_j$ acting on a primary spits out a primary in stead of a descendant.

The subgroup, $SL(2,\mathbf{R})$ is defined by $\mathbb{R}=\mathbb{I}$, $\vec{v}=0$, $f=0$ and generated by $H$, $D$ and $C$. Evidently, the $(t,0)$ slice is an invariant domain of \sl. Using this \sl algebra, one can reorganize the operator content. A \sl primary $\CO$ is defined by requiring $[C,\CO(0,\vec{0})]=0$. Thus all the primaries defined by \eqref{schp} are \sl primaries but not the other way around. The situation is reminiscent of 2D conformal field theory where we have Virasoro primaries as well as \sl primaries and the \sl primaries are called quasi-primaries. We will borrow that nomenclature and call the Schr\"odinger primaries as \textit{primaries} while we name \sl primaries, \textit{quasi-primaries}. Remarkably the notion of quasi-primaries goes through even for a zero charge sector. Henceforth, by $\phi(t)$ (or $\CO(t)$), we will mean the operator $\phi(t,\vec{0})$(or $\CO(t,\vec{0})$).\\

For a \sl invariant field theory, there is a notion of unitarity/reflection positivity, which guarantees that the two point correlator of two operators inserted at imaginary time $-\t$ and $\t$ is positive definite. We will exploit the \sl subgroup of Schr\"odinger group to borrow the notion of \textit{reflection positivity} in Schr\"odinger field theory. 
%
We consider the following states for $\alpha\in\mathbf{R}$ and $\beta\in\mathbf{R}$:
\begin{align}
 \nonumber |\psi_{\alpha}(\t_1)\rangle&=\int d\t\ \left[\delta(\t-\t_1)+2\t_1\alpha^{-1}\delta^{\prime}(\t-\t_1)\right]\CO(\t)|0\rangle\,,\\
\nonumber &=\bigg|\left[\CO(\t_1)-2\t_1\alpha^{-1}(\partial_\t\CO)(\t_1)\right]\bigg\rangle\,.\\
\nonumber |\Psi_{\beta}(\t_1)\rangle&=\int d\t\ \bigg[\left(\frac{Nd}{2\Delta}-\frac{N}{\beta}\right)\delta^{\prime}(\t-\t_1)\phi^\dagger(\t)|0\rangle\\
\nonumber&+ \delta(\t-\t_1)\mathcal{A}^\dagger(\t)|0\rangle\bigg]=\bigg|\left[\frac{N}{\beta}\partial_\t\phi^\dagger-\frac{1}{2}\nabla^2\phi^\dagger\right]\bigg\rangle_{\t=\t_1}\\
\end{align}
where $\CO$ is a quasi-primary,
$\phi^\dagger$ is a primary with charge $-N$
and $\mathcal{A}^\dagger\equiv\left(\frac{Nd}{2\Delta}\partial_\t\phi^\dagger-\frac{1}{2}\nabla^2\phi^\dagger\right)$ is a quasi-primary\footnote{The details of \sl invariant field theory and Schr\"odinger algebra can be found in the appendix~\ref{app1} and~\ref{app2}.}.

To derive the unitarity bound for quasi-primary, we demand that the state $|\psi_\alpha\rangle=|\psi_{\alpha}(1/2)\rangle$\footnote{Technically, the ratio, $\alpha/\t_1$ is dimensionless.  So, in some suitable unit, one can choose $\t_1=1/2$ and vary $\alpha$ in the same unit.} has a positive norm:
\begin{equation}
\langle \psi_\alpha|\psi_\alpha\rangle\geq0\Leftrightarrow \Delta^2+(2\alpha+1)\Delta+\alpha^2\geq 0\,,
\end{equation}
where we have used $\langle\CO(\t_1)\CO(\t_2)\rangle=(\t_2-\t_1)^{-\Delta}$.
\begin{figure}[!ht]
\includegraphics[scale=0.55]{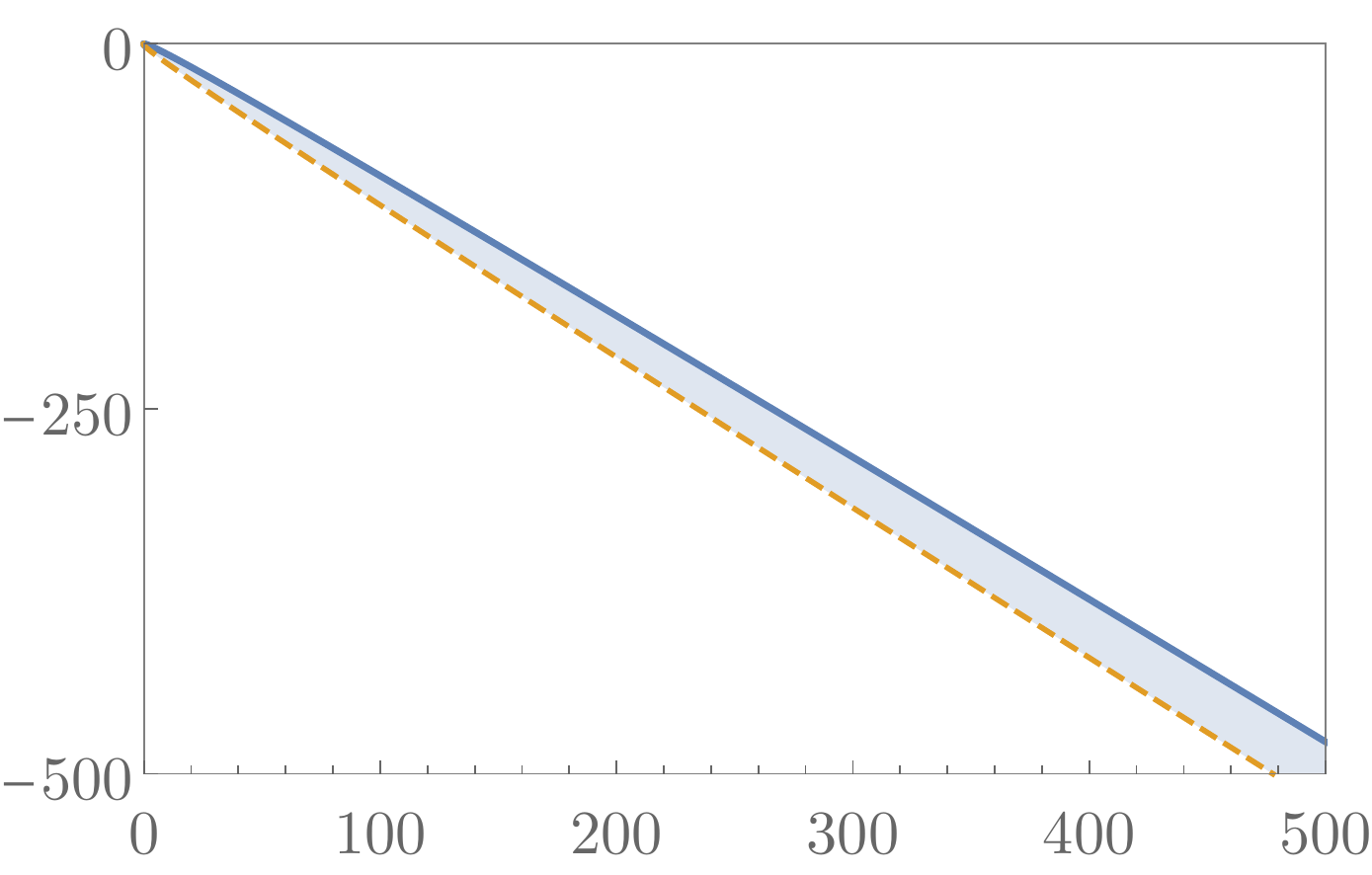}
\caption{Unitarity bound on $(\alpha,\Delta)$ plane: the projection of the region bounded by two curves onto the $Y$ axis excludes $\Delta<0$. The blue thick curve is $\Delta_{+}=-\alpha-\frac{1}{2}+\sqrt{\alpha+\frac{1}{4}}$ while the orange dashed curve is $\Delta_{-}=-\alpha-\frac{1}{2}-\sqrt{\alpha+\frac{1}{4}}$.}
\label{normqp}
\end{figure}
For $\alpha<-\tfrac{1}{4}$ there is no constraint on $\Delta$. For $\alpha\geq-1/4$, the region $\left(\Delta_-,\Delta_+\right)$ is excluded where
\begin{align}
\Delta_{\pm}=(-\alpha-1/2\pm\sqrt{\alpha+1/4})\leq 0.\,.
\end{align}
As we vary $\alpha$, the whole $\Delta<0$ region gets excluded (fig.~\ref{normqp}) 
since, $\Delta_+-\Delta_-=2\sqrt{\alpha+1/4}$. 

Now we will do the same for a primary and consider the norm of the state $|\Psi_\beta\rangle=|\Psi_\beta(1/2)\rangle$,
\begin{equation}\label{ineq2}
\langle\Psi_\beta|\Psi_\beta\rangle\geq0\Leftrightarrow \Delta^2+\Delta(1-\beta  d)+\frac{1}{4} \beta  d (\beta  (d+2)-4) \geq 0
\end{equation}

\begin{figure}[!ht]
\includegraphics[scale=0.55]{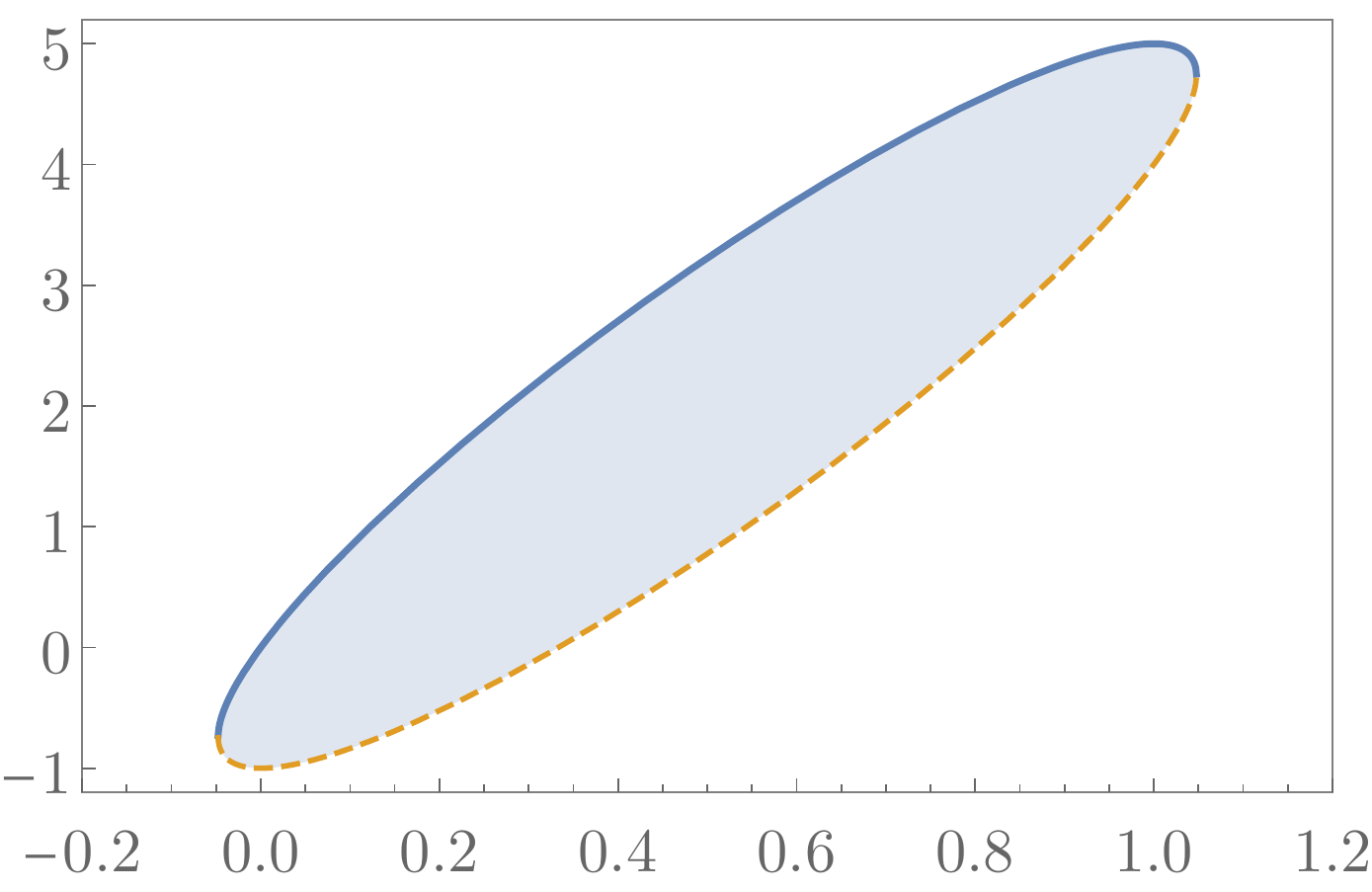}
\caption{Unitarity bound on $(\beta,\Delta)$ plane: the projection of the region bounded by the ellipse onto the Y axis excludes $\Delta\in\left(-1,\frac{d}{2}\right)$. Here $d=10$. The blue thick curve is $\Delta_+= \frac{d\beta-1}{2}+\frac{1}{2}\sqrt{1+2d\beta-2d\beta^2}$ while the orange dashed curve is $\Delta_-= \frac{d\beta-1}{2}-\frac{1}{2}\sqrt{1+2d\beta-2d\beta^2}$. }
\label{normp}
\end{figure}

To find out the norm of $\Psi_\beta$, we have used the two point correlator of primaries, fixed by Schr\"odinger algebra. This is exactly where it becomes important that the actual symmetry group is bigger than \sl and there are $d$ spatial dimensions. The region, excluded (fig.~\ref{normp}) due to \eqref{ineq2} is given by $\left(\Delta_-,\Delta_+\right)$ where $\Delta_\pm= \frac{d\beta-1}{2}\pm\frac{1}{2}\sqrt{1+2d\beta-2d\beta^2}$.

As we vary $\beta$, on the $(\beta,\Delta)$ plane, the excluded region is bounded by an ellipse. This, in turn, excludes $\Delta\in\left(-1,\frac{d}{2}\right)$. Thus we have (recalling a primary is a quasi-primary too and has to satisfy the bound for quasi-primary)
\begin{equation}
\Delta \notin \left(-1,\frac{d}{2}\right) \cup \left(-\infty,0\right) \Rightarrow \Delta \geq \frac{d}{2}.
\end{equation}

The bound is saturated when $\Delta_+=\frac{d}{2}$, which implies that $\beta=1$ and we arrive at the null condition: 
\begin{equation}\label{null}
\langle\Psi_{1}|\Psi_1\rangle =0 \Leftrightarrow N\partial_\t\phi^\dagger-\frac{1}{2}\nabla^2\phi^\dagger=0
\end{equation}
The unitarity bound and the null condition, thus obtained for the charged sector, is consistent with the results in \cite{Nishida:2010tm}. A technical remark is in order: setting $\beta=1$ to begin with, would not suffice to derive the unitarity bound. This is because merely demanding$\langle\Psi_{1}|\Psi_1\rangle\geq0$ would exclude the region $\left(\frac{d}{2}-1,\frac{d}{2}\right)$ only.

For a sector with $N=0$, the unitarity bound becomes the one obtained by using $|\psi_\alpha\rangle$, thus the null condition is achieved when $\Delta_+(\alpha)=0\Rightarrow\alpha=0$. Thus the null condition for neutral sector reads $\partial_\t\CO=0$. The bound in the neutral sector is lower compared to the bound in charged sectors, thus in free Schr\"odinger field theory there's no neutral operator satisfying the bound except the Identity operator. The identity operator by definition creates the neutral vacuum state, has $0$ dimension and trivially time independent. It would be interesting to find an operator besides the identity operator, which saturates the bound or improve the bound for the non-identity operators. One might hope to come up with stronger bound for the neutral sector by considering the norm of the state $A|\psi_\alpha\rangle+B|\Psi_\beta\rangle$, but this is given by $A^2\langle\psi_\alpha|\psi_\alpha\rangle+B^2\langle\Psi_\beta|\Psi_\beta\rangle$ since $|\Psi_\beta\rangle$ is charged whereas $|\psi_\alpha\rangle$ is neutral, leading to $\langle\psi_\alpha|\Psi_\beta\rangle=0$. Now, $A^2\langle\psi_\alpha|\psi_\alpha\rangle+B^2\langle\Psi_\beta|\Psi_\beta\rangle\geq0$ by previous bounds.\\

\paragraph*{\bf Subtleties associated with Null condition \& Non-Renormalization:}
The derivation of the null condition assumes that the only operator that can annihilate the vacuum is the null operator (denoted as $\hat{0}$ henceforth). This is not necessarily true in a non-relativistic set up. For example, the canonical way of quantizing free Schr\"odinger field theory starts with the existence of an operator $\phi$ such that $\phi$ annihilate the vacuum  Thus for $\t>0$, we have $\langle 0|\phi^{\dagger}(0)\phi(\tau)|0\rangle= 0$. But this does not imply that $\phi$ is a null operator. In a theory with anti-particles, $\phi$ can not annihilate the vacuum since its Fourier decomposition consists of several particle annihilation operators and anti-particle creation operators. But a non-relativistic field theory admits a quantization process without having any anti-particle in its spectrum. Thus, non-trivial operators like $\phi$ can have the vacuum state as their kernel. 

To state a generic null condition, we consider the set of operators $\mathcal{S}_N$, defined by
$s_N\in \mathcal{S}_{N}$ iff $|0\rangle \in \ker(s_N)$, $[\hat{N},s_N]=Ns_N.$
The null condition then reads:
\begin{equation}\label{nullm}
N\partial_\t\phi^\dagger-\frac{1}{2}\nabla^2\phi^\dagger \in \mathcal{S}_{-N} \cup \{\hat{0}\}\,.
\end{equation}
We see that, unlike relativistic CFT, the Eq.~\eqref{nullm} can be satisfied at an interacting fixed point. It has, therefore, consequences in terms of anomalous dimension of $\phi$. For example, let us consider a free Schr\"odinger field theory and perturb by an operator of the form $s(x)\phi(x)$ where $s\in \mathcal{S}_{N}$. If the theory flows to another fixed point such that Eq.~\eqref{nullm} holds, the field $\phi$ can not acquire an anomalous dimension. This happens because the null condition~\eqref{nullm} implies that even at the non-trivial fixed point $\phi$ has dimension $\frac{d}{2}$, which equals the dimension at the free fixed point.

The non-renormalization theorem can be utilized in following way: consider a free Schr\"odinger field theory with free elementary fields $\phi_\alpha$  (the ones that appear in Lagrangian at free fixed point) and $[\hat{N},\phi_\alpha]=N_\alpha\phi_\alpha$ with $N_\alpha<0$. We further assume without loss of generality that $\phi^\dagger_1$ has the minimum positive charge given by $-N_{1}>0$. The absence of anti-particles mean $\phi_{\alpha}$ annihilates the vacuum. Now we perturb the theory by adding a classically marginal $s_{-N_{1}}\phi_1+h.c$ term, where $s_{-N}$ carries charge $-N>0$ and annihilates the vacuum. Assuming that the theory flows to a another fixed point invariant under Schr\"odinger symmetry, we can show that the field $\phi_{1}$ does not acquire any anomalous dimension at the non trivial fixed point. 
 
We proceed by observing that all the terms that might get generated due to renormalization group flow preserve $U(1)$. Furthermore, we only look for the operators of the form $s^{\prime}\phi_1$, as they contribute to the equation motion of $\phi_{1}^\dagger$. Now, the $U(1)$ charge conservation guarantees that $s^{\prime}$ has $-N_1$ charge. We need to show that $s^\prime$ annihilates the vacuum. This would not be the case if $s^{\prime}=\phi_{1}^\dagger$, but this operator can not be generated from a classically marginal term. So we are left with the other option which requires having at least two elementary field operators such that their charges add up to $-N_1$. Since $-N_1$ is the least possible positive charge, there exists at least one operator with negative charge and this implies that $s^{\prime} \in\mathcal{S}_{-N_{1}}$ i.e $s^{\prime}$ annihilates the vacuum. Thus the null condition $(N_1\partial_\t+\nabla^2)\phi^\dagger|0\rangle=0$ is always satisfied for the field with the least possible charge and the corresponding field operator does not acquire any anomalous dimension at the nontrivial fixed point. For example, fermions at unitarity is described by two equivalent theories living at a non-trivial Wilson-Fisher fixed point: one in $2+\epsilon$ dimensions, another one in $4-\epsilon$ dimensions. It is easy to verify from \cite{Nishida:2007pj} that both of them conform to the above theorem. The one fermion operator $\psi$ does not acquire anomalous dimension in both $2+\epsilon$ and $4-\epsilon$ dimensions whereas in $4-\epsilon$ dimensions, the two fermion operator $\phi$ does acquire a anomalous dimension, which should be the case since even at tree level the equation motion of $\phi^\dagger$ does not belong to $\mathcal{S}_{-N}$ where
\begin{align}
s_{-N}\in \mathcal{S}_{-N}\quad &\text{iff}\quad |0\rangle \in \ker(s_{-N})\ \&\ [\hat{N},s_{-N}]=-Ns_{-N}\,.
\end{align}

\section{ Non Unitarity in non integer dimensions}\label{nonuni}
The unitarity of a \sl invariant field theory can be defined in non-integer dimensions by analytically continuing the appropriate correlator. Relativistic CFTs in non integer dimensions can have \textit{Evanescent operators}, corresponding to states with negative norm, thus has a non-unitary sector. These operators cease to exist whenever $d$ becomes integer, nonetheless they are present and non-trivial whenever one extends the theory away from integer (spatial) dimensions \cite{Ben,Gracey:2008mf,Hogervorst:2015akt,yao}. Here we consider a free Schr\"odinger field theory in $d+1$ dimensions and show the presence of such operators. In particular, we consider the following set of operators for $n\geq2$,
\begin{equation*}
R_{n}(t,\vec{x}):= \delta^{i_1[j_1}\delta^{|i_2|j_2}\cdots\delta^{|i_n|j_n]} :M_{i_1j_1} M_{i_2j_2}\cdots M_{i_nj_n}:,
\end{equation*}
where $M_{ij}=\partial_i\partial_j\phi (t,\vec{x})$, $\phi$ is a primary operator with dimension $d/2$ and all the $j$ indices are anti-symmetrized.  For example, $R_{2}(t,\vec{x}):= (\delta^{i_1j_1}\delta^{i_2j_2}-\delta^{i_1j_2}\delta^{i_2j_1}):M_{i_1j_1} M_{i_2j_2}:$. For integer $d<n$ dimensions, at least one of the indices has to repeat itself, thus the operator becomes trivially zero. For non integer $d<n-1$ and for $d\geq n$, the operators are indeed nontrivial. 

The operator $R_2$ produces a negative norm state (we are using the notion of \textit{state} borrowed from \sl invariant field theory, as explained in the appendix~\ref{app1}) in a theory living on $(1-\epsilon)+1$ dimensions with $1>\epsilon>0$. The norm of $R_2$ is given by
\begin{align}
\big \langle R_2(-\tfrac{1}{2},0) R_2^{\dagger}(\tfrac{1}{2},0)\big \rangle = \# \left(d+2\right)\left(d+1\right)d\left(d-1\right)
\end{align}
where $\#$ is a positive number, determined by the two point correlator of $\phi$ and number of independent ways to contract. Here we have also set $\tau=\frac{1}{2}$ without any loss of generality. As expected, the norm becomes zero as $d=0,1$. The norm is negative when $0<d<1$. Similarly, we find that
\begin{align}
\nonumber \big \langle & R_3(-\tfrac{1}{2},0) R_3^{\dagger}(\tfrac{1}{2},0)\big \rangle = \# \left(d+2\right)\left(d+1\right)d^2\left(d-1\right)\left(d-2\right)\\
\nonumber \big \langle & R_4(-\tfrac{1}{2},0) R_4^{\dagger}(\tfrac{1}{2},0)\big \rangle \\
\nonumber&= \# \left(d+2\right)\left(d+1\right)d^2\left(d-1\right)^2\left(d-2\right)\left(d-3\right)\\
\nonumber \big \langle & R_5(-\tfrac{1}{2},0) R_5^{\dagger}(\tfrac{1}{2},0)\big \rangle \\
\nonumber&= \# \left(d+2\right)\left(d+1\right)d^2\left(d-1\right)^2\left(d-2\right)^2\left(d-3\right)\left(d-4\right)
\end{align}
where $\#$ is a positive number, determined by the two point correlator of $\phi$ and number of independent ways to contract. In general, the operator $R_n$ produces a negative norm state:
\begin{equation*}
\big \langle R_n(-\tfrac{1}{2},0) R_n^{\dagger}(\tfrac{1}{2},0)\big \rangle = \# \left(d-n+3\right)_{n}\prod_{j=0}^{n-1} \left(d-j\right)\,,
\end{equation*}
where $\left(d-n+3\right)_{n}=\frac{\Gamma\left(d+3\right)}{\Gamma\left(d-n+3\right)}$ is the Pochhammer symbol. The norm becomes negative when $(n-2)<d<(n-1)$.

\begin{figure}[!ht]
\includegraphics[scale=0.18]{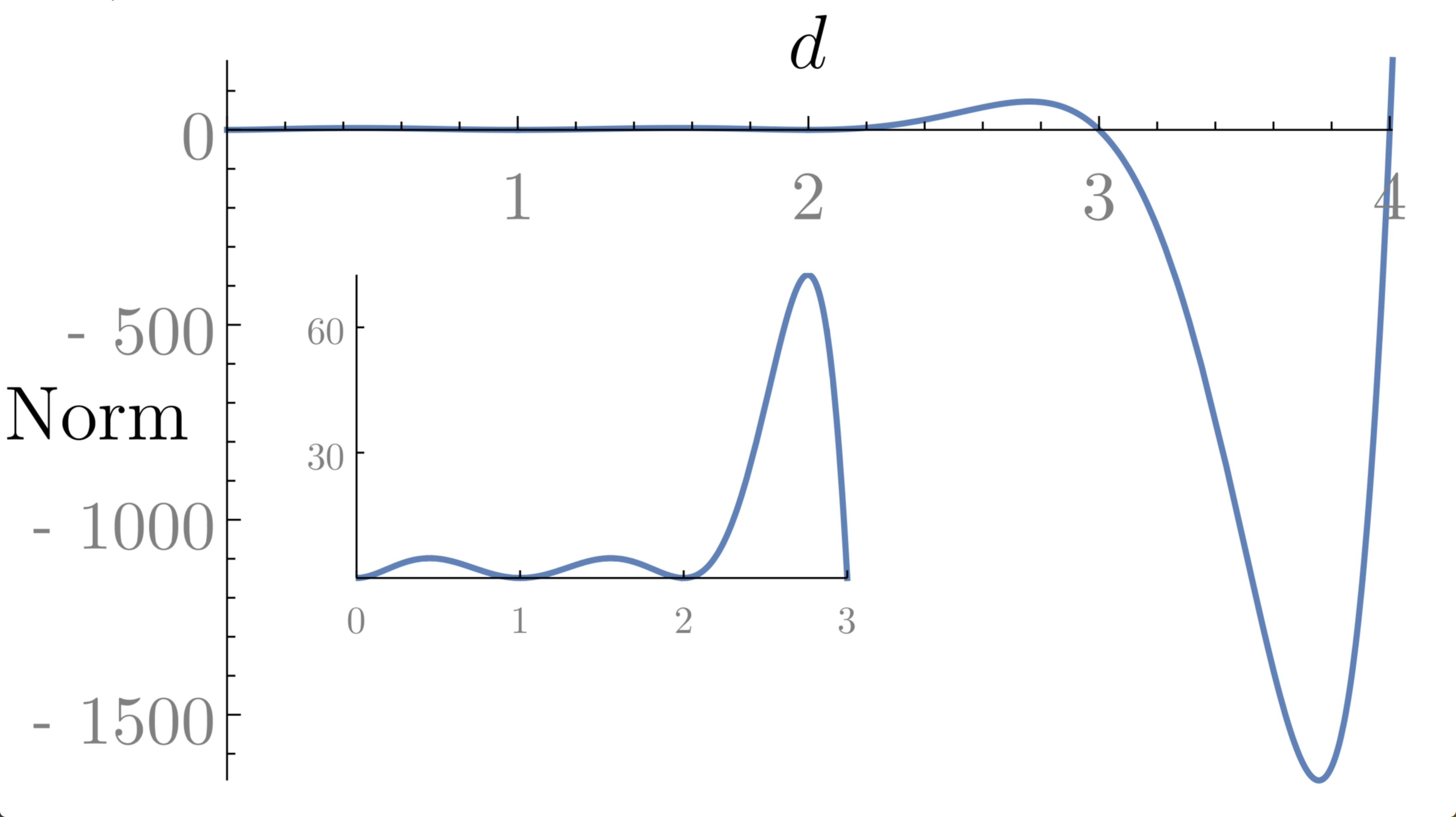}
\caption{The norm of $R_5$ as a function of $d$. This becomes negative for $3<d<4$. The zoomed in version shows that the norm is non negative when $d\in (0,3)$ and becomes zero iff $d$ is an integer below $5$ i.e. $d=0,1,2,3,4$.}
\label{norm}
\end{figure}

These negative norm states are robust and do survive at the Wilson-Fisher fixed point as long as the fixed point can be reached perturbatively i.e. $\epsilon<<1$\cite{Hogervorst:2015akt}.\\

\section{Universality}\label{universal}
$(0+1)$-D \sl invariant field theory has universal features, irrespective of the details of the theory. They come out as a natural consequence of \sl invariance and crossing symmetry. Schr\"odinger field theory, by virtue of being \sl invariant theory as well, inherits these universal features, specially in the neutral sector. Following \cite{Qiao:2017xif}, we consider the 4 point correlator of four Hermitian operator $\langle \CO(0)\CO(\t)\CO(1)\CO(\infty)\rangle= \t^{-\Delta_\CO}\mathcal{G}(\t)$, where, from the \sl algebra, it follows that
\begin{align}
\mathcal{G}(\t)&=\int_0^{\infty}d\Delta p(\Delta) G_{\Delta}(\t)\,,\\
 G_{\Delta}(\t)&=\t^{\frac{\Delta}{2}}\ {}_2F_{1}\left(\frac{\Delta}{2},\frac{\Delta}{2},\Delta,\t\right)
\end{align}
where $p(\Delta)$ is the weighted spectral density and given by $|c_{\CO\CO \Delta}|^2\rho(\Delta)$. Here $\rho(\Delta)$ is the density of quasi-primaries at $\Delta$, $c_{\CO\CO\Delta}$ is the three point coefficient and $G_{\Delta}$ is \sl block. We refer to the \eqref{tpc} in appendix~\ref{app1} for the generic form of three point correlator. 
The convergence of this integral for finite $\t$ holds true for the same reason it holds true in $(0+1)$-D conformal field theory. 
%
Now, as $\t\to 1$, the operator product expansion (OPE) of $\CO(\t)\CO(1)$ is dominated by the contribution from the identity operator, thus we have 
\begin{align}
\mathcal{G}(1-\t)\sim \t^{-\Delta_\CO}\,,\quad \t\to 0\,.
\end{align}
Using the fact, 
\begin{align}
G_{\Delta}(1-\t)\simeq2^{\Delta}\sqrt{\frac{\Delta}{2\pi}}K_0\left(\sqrt{\t}\Delta\right)\,,
\end{align}
one can obtain~\cite{Qiao:2017xif}:
\begin{equation}\label{asymp}
p(\Delta)\underset{\Delta\to\infty}{\sim} 2^{-\Delta}\sqrt{\frac{2\pi}{\Delta}} \frac{4^{1-\Delta_\CO}}{\Gamma(\Delta_\CO)^2}\Delta^{2\Delta_\CO-1}
\end{equation}
The difference in factors of $2$, as compared to Ref.~\cite{Qiao:2017xif}, is coming from the definition of the dilatation operator and $\Delta$ in Schr\"odinger field theory. We remark that Schr\"odinger group has $U(1)$ subgroup, invariance under which implies that each of the operator $\CO$ carries zero charge under $U(1)$. Thus, we are in fact probing the neutral sector, where one can not define the notion of  Schr\"odinger primary. 
Furthermore, the non-zero asymptotics of $p(\Delta)$ in Eq.~\eqref{asymp} directly implies that there has to be infinite number of quasi-primaries.\\

\paragraph*{\textbf{Infinite number of quasi-primaries:}}
One can prove the existence of infinite number of quasi-primaries  in the $\CO\CO\to\CO\CO$ OPE channel using the crossing symmetry as well. It might seem that the existence of infinite number of quasi-primaries are trivial as in the charged sector, operators that are some number of spatial derivatives acting on a primary do appear and they can be written down as a linear combination of \sl descendants and quasi-primaries. But, here we consider $\langle \CO(0)\CO(\t)\CO(1)\CO(\infty)\rangle$ and all the operators lie at $\vec{x}=0$. As a result, the operators that appear in the OPE are not of the form of some spatial derivative acting on a primary. 

The proof goes by noting that the crossing symmetry implies 
\begin{align}\label{cross}
(1-\t)^{\Delta_\CO}\mathcal{G}(\t)=\t^{\Delta_\CO}\mathcal{G}(1-\t)
\end{align}
where $\Delta_\CO$ is the dimension of the operator $\CO$. 

As $\t\to0$, the leading contribution to the left hand side of \eqref{cross} comes from identity, i.e. we have $(1-\t)^{\Delta_\CO}\mathcal{G}(\t)=1+\cdots$. If we look at the right hand side in terms of blocks, we realize that each $G_{\Delta}(1-\t)$ goes like $\log(\t)$, thus the each term in the block decomposition of $\t^{\Delta_\CO}\mathcal{G}(1-\t)$ behaves like $ \t^{\Delta_\CO}\log(\t)$, which goes to $0$ as $\t\to0$. If we have finite number of quasi-primaries, since each of the summands goes to $0$, we could never have \eqref{cross} satisfied in $\t\to0$ limit. This proves the existence of infinite number of quasi-primaries which are not descendants of a primary. Similar argument works for \sl primaries of conformal field theories as well. This line of argument has first appeared in \cite{Rattazzi:2008pe} (See also\cite{Simmons-Duffin:2016gjk}).\\

\paragraph*{\textbf{Analyticity of Three point function:}}
The three point function of Schr\"odinger primaries are given by 
\begin{align}
\nonumber&\langle \phi_1(\vec{x}_1,t_1)\phi_2(\vec{x}_2,t_2)\phi_3(\vec{x}_3,t_3) \rangle\\ 
&= \exp\left[-N_1\frac{|\vec{x}_{13}|^2}{\t_{13}}-N_2\frac{|\vec{x}_{23}|^2}{\t_{23}}\right]\bigg(\prod_{i<j}\t_{ij}^{\frac{\Delta_k-\Delta_i-\Delta_j}{2}}\bigg)F(v_{123})
\end{align}
where $v_{123}=\frac{1}{2}\left(\frac{|\vec{x}_{23}|^2}{\t_{23}}+\frac{|\vec{x}_{12}|^2}{\t_{12}}-\frac{|\vec{x}_{13}|^2}{\t_{13}}\right)$ and $\vec{x}_{ij}=\vec{x}_j-\vec{x}_i$, $\t_{ij}=\t_j-\t_i$. $F$ is a model dependent function and zero if $\sum_iN_i\neq0$, where $N_i$ is the charge carried by $\phi_i$. By translation invariance, we can set $\t_3=0$ and $\vec{x}_3=0$. As $\phi(\vec{0},\t)$ is a quasi-primary, upon setting $\vec{x}_1=\vec{x}_2=\vec{x}_3=0$, we immediately obtain that $F(0)$ is a finite number and given by three point coefficient $c_{123}$.

For simplicity, let us work in $d=1$ dimension and set $x_1=x_3=0$ and investigate the behavior of $F$ as a function of $x_2$. The \sl algebra guarantees that $F$ is infinitely differentiable at $x_2=0$. This follows from noting that 
\begin{align}
\nonumber &\langle \phi_1(0,t_1)\phi_2(x_2,t_2)\phi_3(0,0)\rangle \\
&= \sum_{\beta}\frac{1}{\beta!}x_2^{\beta} \langle\phi_1(0,t_1)\partial_{x_2}^\beta\phi_2(0,t_2)\phi_3(0,0)\rangle
\end{align}
and finiteness of $\langle\phi_1(0,t_1)\partial_{x_2}^\beta\phi_2(0,t_2)\phi_3(0,0)\rangle$. The finiteness follows from finiteness of norm and the fact that $\partial_{x_2}^\beta\phi_2(0,t_2)$ can be written down as a linear combination of quasi-primaries and descendants of quasi-primary. For example, we list out the first two quasi-primaries (in $d$ spatial dimensions), given $\phi$ is a primary (which is trivially a quasi-primary too): 
\begin{align}
\nonumber \mathcal{B}^{(1)}_i\equiv\partial_{i}\phi\,,\quad \mathcal{B}^{(2)}\equiv\frac{Nd}{2\Delta}\partial_\t\phi+\frac{1}{2}\nabla^2\phi\,.
\end{align}

\section{Discussion \& Outlook}
We have shown that the features of $(0+1)$-D conformal field theory is inherited by the Schr\"odinger field theory. \sl algebra can be leveraged to derive the unitary bounds and null conditions, to prove the convergence of operator product expansion in the kinematic limit, where all the operators are inserted at same $\vec{x}$, but at different times. Moreover, if we consider the four point correlator of  Schr\"odinger primaries with all but one inserted on $(\t,\vec{0})$ slice and one operator inserted at some different $\vec{x}\neq\vec{0}$, we can still prove the OPE convergence by using \sl invariance. This happens because $\vec{x}$ dependence of the four point correlator is simply given by $\exp\left[\tfrac{N|\vec{x}|^2}{2\t}\right]$ where $N<0$ is the charge of the operator. The use of \sl reveals the universal behavior of the weighted spectral density function and the existence of infinitely many quasi-primaries. We emphasize the salient role of \sl in this context, as concept of Schr\"odinger primaries and descendants break down in the neutral sector. Moreover, one can easily deduce the analyticity of three point co-efficient function as a consequence of \sl. It worths mentioning that the usual oscillator picture also relies on \sl algebra in hindsight. In fact, the state-operator correspondence using the oscillator picture work beyond the primary operator: for every quasi-primary operator, one can define a state $|\CO\rangle=e^{-H}\CO|0\rangle$ such that $\left(H+C\right)|\CO\rangle=\Delta|\CO\rangle$, where $H+C$ can be interpreted as a Hamiltonian for the same system under harmonic trap. 

The use of \sl algebra provides us with a neat way to define the Euclidean Schr\"odinger theory. We refer to the appendix~\ref{app4} for more details. This justifies the Wick rotation done in \cite{Pal:2017ntk} to evaluate the heat kernel and the Weyl anomaly. Moreover, the use of Euclidean Schr\"odinger operator in \cite{Auzzi:2016lxb,Solodukhin:2009sk} comes under question in this light as the correlator obtained from the heat kernel of such operators do not satisfy the constraint coming from \sl algebra. In this connection, it deserves a remark that the notion of parity ($\t \to-\t$) and time reversal ($\t \to-\t$ with charge conjugation) is subtle in $(0+1)$-D conformal field theory and the same subtlety is also present in Schr\"odinger field theories (the details have been relegated to the appendix~\ref{app3}). If one can consistently impose parity invariance beyond $(\t,\vec{0})$ slice, 
such theories should have anti-particles and are suspected to have a non-zero entanglement entropy in the vacuum in contrast with its cousin where anti-particles are absent. One also wonders about presence of Weyl anomalies in such parity invariant theories on coupling to a non trivial curved background in the same spirit of \cite{Arav:2016xjc,Auzzi:2016lrq,Pal:2016rpz,Jensen:2014hqa,Jensen:2014aia,Geracie:2014nka}.

The most important take home message is that bootstrapping the Schr\"odinger field theory on $(\t,0)$ slice exactly amounts to bootstrapping $0+1$ D conformal field theory. Thus one can extend the analysis for 4 point correlator of operators with different dimensions, not necessarily the Hermitian ones with an aim to use \sl bootstrap \cite{Hogervorst:2017sfd,Mazac:2016qev,Rychkov:2017tpc,El-Showk:2016mxr} program to derive useful constraints for Schr\"odinger field theories. Furthermore, the four point correlator $\langle \CO(0)\CO(t)\CO(1)\CO(\infty)\rangle$ is analytic in complex $t$ domain. One might hope to gain more mileage for $(0+1)$-D conformal field theory as well as the Schr\"odinger field theory using analyticity in the complex plane \cite{Hogervorst:2017sfd,Mazac:2016qev}. In fact, if one is interested in knowing the spectra of the dilatation operator, then bootstrapping on $(\t,0)$ slice is sufficient as well. Should one consider a four point correlator of operators $\CO_i$ inserted at different $\vec{x}$, the OPE would have operators $\left[P_{i_1},[P_{i_2},\cdots[P_{i_n},\CO_{k}]]\right]$ while on the $(\t,0)$ slice, we would only have $\CO_k$ operator. But the dimension of $\left[P_{i_1},[P_{i_2},\cdots[P_{i_n},\CO_{k}]]\right]$ is completely fixed by $\CO_k$. This feature elucidates why it is sufficient to bootstrap on the $(\t,\vec{0})$ slice to know the spectra of the dilatation operator. Similar argument applies for knowing the OPE coefficients.

On a different note, the operator basis for the heavy quark effective field theory (HQEFT), non relativistic QED/QCD \cite{Manohar:1997qy} can be organized according to the representation of the Schr\"odinger algebra (or of \sl algebra) like it is done for the Standard Model effective field theory  \cite{Henning:2015daa,Henning:2015alf,Henning:2017fpj}. 
As the operators appearing in the Lagrangian of HQEFT are necessarily neutral, the concept of quasi-primary is quintessential in that context as reported on a separate paper \cite{Kobach:2018nmt} with an application towards construction of an operator basis\cite{Hill:2012rh,Kobach:2017xkw,Gunawardana:2017zix} for heavy particle effective field theory.

There are further questions which requires more attention. Fermions at unitarity \cite{Nishida:2007pj} is described by a nontrivial fixed point in $4-\epsilon$ dimensions, it is important to investigate whether there is any imprint of non-unitarity in the physics of that fixed point. A step towards this would be to find out whether heavy enough operators acquire complex anomalous dimension at WF fixed point. For a relativistic scenario, this has been done in \cite{Hogervorst:2015akt}.  It is also worthwhile to investigate whether \sl constrains the properties of a thermal Schr\"odinger field theory \cite{Jensen:2014ama}. At $0$ temperature, one can calculate all the  correlators using the OPE coefficients. For $T>0$, the OPE is expected to hold true for time $|t|<<\frac{\hbar}{k_BT}$\cite{Katz:2014rla}. Thus using the \sl algebra, it seems possible to obtain sum rules involving conductivities as done in \cite{Katz:2014rla,Lucas:2016fju,Lucas:2017dqa}, particularly for CFTs. Furthermore, the idea presented here is extendable to the theories invariant under a symmetry group which contains \sl as subgroup. The natural question is to ask whether the generalized $z$ ($z\neq2$) group can have a bound. It is shown\cite{Balasubramanian:2008dm} that the algebra does not close with the special conformal generator $C$, if one has the particle number symmetry generator $\hat{N}$. Thus \sl subgroup is absent and they can not be realized with finite dimensional basis of operators \cite{Grinstein:2018xwp}. Nonetheless, if one does not have the $U(1)$ associated with particle number symmetry, the algebra closes with $C$ and it does have a \sl piece, so similar analysis can be done for field theories invariant under such group. For sake of completeness, we write down the algebra so that \sl becomes manifest:
\begin{align}
\nonumber [D,C]&=-2\imath C\,,[D,H]=2\imath H\,,[H,C]=-\imath D\\
\nonumber [D,P_i]&=\imath\alpha P_i\,,[D,K_i]=2\imath(\alpha-\alpha^{-1})K_i\,,[H,K_i]=-\imath P_i\\
\nonumber [H,P_i]&=[P_i,P_j]=[K_i,P_j]= 0
\end{align}
where $\alpha=\frac{1}{z}$. 
The commutation relations of these with the generators of rotation group are the usual ones. Last but not the least, in $1+1$ dimensions, \sl algebra gets extended to infinite Virasoro algebra. One can then introduce Virasoro conformal blocks and one has more analytical control over such theories. One wonders whether there exists any such extension for the Schr\"odinger algebra. If exists, it would imply the possibility of borrowing the arsenal of Virasoro algebra.

\begin{acknowledgments}
I acknowledge the support provided by the US Department of Energy (DOE) under cooperative research agreement DE-SC0009919. It is a pleasure to thank Ken Intriligator, Kristan Jensen, William Krempa, John McGreevy, Slava Rychkov for insightful comments and discussions. I thank Slava Rychkov to point out that nonzero asymptotics of weighted OPE coefficients directly implies existence of infinitely many quasi-primaries. I acknowledge fruitful discussions with Andrew Kobach, Aneesh Manohar and Jaewon Song regarding using Schr\"odinger algebra to construct operator basis for HQET/NRQCD. I also thank Peter Stoffer and Brato Chakrabarti for help with Mathematica.
\end{acknowledgments}

\appendix
\section{\sl invariant field theory}\label{app1}
The \sl group is generated by the three generators $H,D,C$ satisfying the following algebra: 
\begin{align}
[D,H]= 2 H\,,\ [D,C]=-2C\,,\ [H,C]=D.
\end{align}
The spectrum of $D$ is real and physically represents the dimension. $H$ raises the dimension and $C$ lowers the same. The highest weight representation is the one, annihilated by $C$. A nice and brief exposition of \sl invariant field theory can be found in the appendix of \cite{Qiao:2017xif}. Here we discuss them for the sake of completeness and make the paper self-contained.

An \sl invariant field theory is defined on a one dimensional manifold, parameterized by $\tau$ (say \textit{time}), where \sl group acts on the co-ordinate $\t$ in following way:
\begin{align}
\t \mapsto \frac{a\t+b}{c\t+d}\,,\quad \text{where}\ (ad-bc)=1
\end{align}
In terms of the generators above, the $H$ generates a time translation, $C$ generates a special conformal transformation and $D$ generates a scale transformation. The theory has a privileged class of operators (the highest weight representation) $\CO_\alpha$ satisfying
\begin{align}
[C,\CO_\alpha(0)]=0
\end{align}
and carrying dimension $\Delta_\alpha$ i.e $[D,\CO_\alpha(0)]=\Delta_\alpha \CO_\alpha(0)$. These are called \sl primaries. They have following commutators with \sl generators:
\begin{align}
[D,\CO_\alpha(\t)]&=\left(2\tau\partial_\tau+\Delta_\alpha\right)\CO_\alpha(\t)\,,\\
[C,\CO_\alpha(\t)]&=\left(- \t^2\partial_\t-\t\Delta_\alpha\right)\CO_\alpha(\t)\,,\\
[H,\CO_\alpha(\t)]&=\partial_\t\CO_\alpha(\t)\,.
\end{align}

It follows that under a finite \sl transformation, \sl primaries transform as 
\begin{align}
\CO(\t)\mapsto \left[(c\t+d)^{-2}\right]^{\tfrac{\Delta}{2}} \CO(\t^\prime)
\end{align}
\footnote{The presence of factor of $\frac{1}{2}$ in the weight, as compared to the weight, noted in \cite{Qiao:2017xif} is due to the presence of extra $2$ with $\tau\partial_\t$ in the expression for $[D,\CO_\alpha(\t)]$. This is done, in hindsight, to make the notation consistent with the dilatation operator in Schr\"odinger field theory i.e the dilatation operator acting on a Schr\"odinger primary at $\vec{x}=0$ becomes the dilatation operator acting on a \sl primary.}

The correlators of the form $\langle\CO(\t_1)\CO(\t_2)\CO(\t_3)\cdots\CO(\t_n)\rangle$ will be of our ``primary" interest. As \sl preserves the ordering of time, the two distinct ordering of $\t_i$ need not be related to each other. In what follows, we will be assuming $0<\t_1<\t_2<\t_3<\cdots<\t_n$. The unitarity for a \sl invariant theory is defined via existence of an anti-linear conjugation map taking $\CO\mapsto\CO^{\dagger}$ such that following relations hold:
\begin{enumerate}
\item \textbf{Time reversal}: 
\begin{align}
\nonumber \langle&\CO^\dagger(-\t_n)\cdots\CO^\dagger(-\t_3)\CO^\dagger(-\t_2)\CO^\dagger(-\t_1)\rangle\\
&=\langle\CO(\t_1)\CO(\t_2)\CO(\t_3)\cdots\CO(\t_n)\rangle^*
\end{align}
\item \textbf{Reflection positivity}: One can define a state $|\Psi\rangle$
\begin{align}
|\psi\rangle=\int \left(\prod d\t_i\right)\ g(\t_1,\t_2,\cdots, \t_n)\ |\CO(\t_1)\CO(\t_2)\cdots\CO(\t_n)|0\rangle
\end{align}
with 
\begin{align}\label{uni}
\langle\Psi|\Psi\rangle\geq 0
\end{align}
In terms of correlator this reads:
\begin{align}
\nonumber \int &\left(\prod d\t^{\prime}_i\right)\ \int \left(\prod d\t_i\right) \tilde{g}(\t^\prime_1,\t^\prime_2,\cdots,\t^\prime_n) g(\t_1,\t_2,\cdots, \t_n)\\
\nonumber&\qquad \langle\CO^\dagger(\t^\prime_n)\cdots\CO^\dagger(\t^\prime_2)\CO^\dagger(\t^\prime_1) \CO(\t_1)\CO(\t_2)\cdots\CO(\t_n)\rangle\geq 0
\end{align}
where $\tilde{g}(\t)=g(-\t)^*$. Here $g$ is an arbitrary function or distribution having support away from coincident points to avoid singularity.
\end{enumerate}

The \sl algebra fixes the functional form of the two point and the three point correlator. One can choose a Hermitian basis of operators $\CO=\CO^{\dagger}$ such that
\begin{align}
\langle &\CO_\alpha(\t_1)\CO_\beta(\t_2)\rangle= \frac{\delta_{\alpha\beta}}{(\t_2-\t_1)^{\Delta}}\\
\nonumber \langle &\CO_1(\t_1)\CO_2(\t_2)\CO_3(\t_3)\rangle\\
\label{tpc}
&= \frac{c_{123}}{(\t_2-\t_1)^{\frac{\Delta_2+\Delta_1-\Delta_3}{2}}(\t_{3}-\t_2)^{\frac{\Delta_3+\Delta_2-\Delta_1}{2}}(\t_{3}-\t_{1})^{\frac{\Delta_3+\Delta_1-\Delta_2}{2}}}
\end{align}
Time reversal symmetry guarantees that $c_{\gamma\beta\alpha}=c_{\alpha\beta\gamma}^{*}$. In general, $c_{\alpha\beta\gamma}$ can be complex numbers. All the higher point correlator can be obtained using the operator product expansion, which reads:
\begin{align}
\CO_1(\t_1)\CO_2(\t_2)= \sum_\alpha c_{12\alpha}\frac{1}{(\t_2-\t_1)^{\frac{\Delta_1+\Delta_2-\Delta_\alpha}{2}}}\left[\CO_\alpha+\cdots\right]
\end{align}
where $\cdots$ are contributions coming from \sl descendants and fixed by \sl invariance. Thus the knowledge of the spectrum of $D$ i.e set of \sl primaries and three point coefficient $c_{\alpha\beta\gamma}$ amounts to a complete knowledge of all the correlators. 

\section{Schr\"odinger algebra, Primaries and Quasi-primaries}\label{app2}

Here we provide a detailed account of Schr\"odinger algebra. We expound the concepts of primaries and quasi-primaries. The Schr\"odinger group acts on space-time as follows\cite{Hagen:1972pd,Niederer:1972zz,Mehen:1999nd,Henkel:2003pu}:
\begin{align}
t\mapsto \frac{at+b}{ct+d}\,, \quad \vec{r}\mapsto \frac{\mathbb{R}\vec{r}+\vec{v}t+f}{ct+d}
\end{align}
where $ad-bc=1$, $\mathbb{R}$ is a d dimensional rotation matrix, $\vec{v}$ denotes the Galilean boost and $f$ is a spatial translation. For the sector with non-zero charge, the representation is built by translating all the operators to the origin and considering the little group generated by dilatation operator $D$, Galilean boost generator $K_i$, and special conformal transformation generator $C$. The highest weight states ($\phi_\alpha$) are annihilated by $C$ and $K_i$ i.e. 
\begin{align}\label{schp}
[C,\phi_\alpha(0,\vec{0})]=0\,,\quad [K_i,\phi_\alpha(0,\vec{0})]=0\,.
\end{align}
These are called primary operators. The commutators with $D$ and particle number symmetry generator $\hat{N}$ dictate the charge and the dimension of these operators $\phi_\alpha$:
\begin{align}
[D,\phi_\alpha(0,\vec{0})]=\imath\Delta_\alpha\phi(0,\vec{0})\,,\ [\hat{N},\phi_\alpha(0,\vec{0})]=N_\alpha\phi(0,\vec{0})
\end{align}
The time and space translation generator $H$ and $P$ create descendant operators by acting upon primary operators, consequently raising the dimension by $2$ and $1$ respectively. It deserves mention that the concept of primaries and descendants breaks down within the neutral sector. Since $K_i$ and $P_j$ commute in this sector, $P_j$ acting on a primary spits out a primary in stead of a descendant.

The subgroup defined by $\mathbb{R}=\mathbb{I}$, $\vec{v}=0$, $f=0$ is generated by $H$, $D$ and $C$ and is in fact $SL(2,\mathbf{R})$. This acts as follows:
\begin{align}
t\mapsto \frac{at+b}{ct+d}\,, \quad \vec{r}\mapsto \frac{\vec{r}}{ct+d}\,,\quad (ad-bc)=1\,.
\end{align}
It becomes evident that the $(t,0)$ slice is a invariant domain of \sl. Now one can reorganize the operator content using \sl algebra. A \sl primary $\CO$ is defined by requiring $[C,\CO(0,\vec{0})]=0$. Thus all the primaries defined by \eqref{schp} are \sl primaries but not the other way around. As mentioned in the main text, the situation is reminiscent of 2D conformal field theory where we have Virasoro primaries as well as \sl primaries and the \sl primaries are called quasi-primaries. We have borrowed that nomenclature and called the Schr\"odinger primaries as \textit{primaries} and \sl primaries as \textit{quasi-primaries}. We emphasize that the notion of quasi-primaries goes through even for the neutral sector. The commutator of quasi-primary $\CO(t)$ [for notational simplicity, $\phi(t)$ (or $\CO(t)$) implies the operator $\phi(t,\vec{0})$(or $\CO(t,\vec{0})$)] with the generators $H$, $D$, $C$ are given by\cite{Goldberger:2014hca}: 
\begin{align}
[H,\CO(t)]&=-\imath\partial_t\CO(t)\,,\\
[D,\CO(t)]&=\imath(2t\partial_t+\Delta)\CO(t)\,,\\
[C,\CO(t)]&=(-\imath t^2\partial_t-\imath t\Delta)\CO(t)\,.
\end{align}
This follows from $[C,\CO(0,\vec{0})]=0$ and 
\begin{align}
[D,H]= 2 \imath H\,,\ [D,C]=-2\imath C\,,\ [H,C]=-\imath D.
\end{align}

In terms of Euclidean time $\tau=\imath t$, and $D^{\prime}=-\imath D$, we have
\begin{align}
[D^{\prime},H]&= 2 H\,,\ [D^\prime,C]=-2C\,,\ [H,C]=D^\prime\,.
\end{align}
and
\begin{align}
[D^\prime,\CO(\t)]&=\left(2\tau\partial_\tau+\Delta\right)\CO(\t)\,,\\
[C,\CO(\t)]&=\left(- \t^2\partial_\t-\t\Delta\right)\CO(\t)\,,\\
[H,\CO(\t)]&=\partial_\t\CO(\t)\,.
\end{align}

Thus we have a \sl invariant theory defined on $(\tau,\vec{0})$ slice. \sl acts on $\t$ in the usual manner:
\begin{align}
\tau\mapsto \frac{a\tau+b}{c\t+d}\,,\quad (ad-bc)=1\,.
\end{align}

We note that if $\phi$ is a primary operator, then
\begin{align}
\mathcal{A}\equiv-\left(\frac{Nd}{2\Delta}\partial_\t\phi+\frac{1}{2}\nabla^2\phi\right)\,,\ \mathcal{A}^{\dagger}\equiv\frac{Nd}{2\Delta}(\partial_\t\phi^\dagger)-\frac{1}{2}\nabla^2\phi^\dagger
\end{align}
are  quasi-primaries but not primaries unless $\Delta=\frac{d}{2}$. This follows from the commutation relations \cite{Goldberger:2014hca}:
\begin{align}
\nonumber [C,\phi(\t,\vec{x})]&=\left(- \t^2\partial_\t-\t\Delta-\t\vec{x}\cdot\vec{\nabla}+\frac{N|\vec{x}|^2}{2}\right)\phi\\
\nonumber [K_j,\phi(\t,\vec{x})]&=\left(-\t\partial_j+Nx_j\right)\phi
\end{align}
The operators $\mathcal{A}$ and $\mathcal{A}^\dagger$ played a crucial role in proving the unitarity bound. In fact, $\mathcal{A}^\dagger$ annihilates the vacuum when $\Delta=\frac{d}{2}$ and at free fixed point, this is precisely the null operator.\\

\section{Time reversal and Parity}\label{app3}
The notion of time reversal and parity is subtle in $0+1$-D conformal field theories. The same subtlety is inherited by the Schr\"dinger field theory. Both of the symmetries come with a transformation of the form $\t\to -\t$, but the time reversal is realized as an anti-unitary operator acting on the fields whereas the parity does not involve any complex conjugation. To be precise, time reversal symmetry guarantees that 
\begin{align}
\nonumber \langle&\CO^\dagger(-\t_n)\cdots\CO^\dagger(-\t_3)\CO^\dagger(-\t_2)\CO^\dagger(-\t_1)\rangle\\
&=\langle\CO(\t_1)\CO(\t_2)\CO(\t_3)\cdots\CO(\t_n)\rangle^*
\end{align}
while the parity invariance implies that
\begin{align}
\nonumber \langle&\CO(-\t_n)\cdots\CO(-\t_3)\CO(-\t_2)\CO(-\t_1)\rangle\\
&=(-1)^{np}\langle\CO(\t_1)\CO(\t_2)\CO(\t_3)\cdots\CO(\t_n)\rangle
\end{align}
where $p\in\{0,1\}$ is the parity of the \sl primary operator $\CO$. 

The three point correlators, as pointed out earlier, are given by
\begin{align}
\nonumber \langle &\CO_1(\t_1)\CO_2(\t_2)\CO_3^{\dagger}(\t_3)\rangle\\
\label{tpc}
&= \frac{C_{\CO_1\CO_2\CO_3}}{(\t_2-\t_1)^{\frac{\Delta_2+\Delta_1-\Delta_3}{2}}(\t_{3}-\t_2)^{\frac{\Delta_3+\Delta_2-\Delta_1}{2}}(\t_{3}-\t_{1})^{\frac{\Delta_3+\Delta_1-\Delta_2}{2}}}
\end{align}

Now the time reversal implies that $C_{\CO_1\CO_2\CO_3}=C_{\CO_2^{\dagger}\CO_1^{\dagger}\CO_3^{\dagger}}^{*}$. Since cyclic ordering is preserved by \sl, we have $C_{\CO_2\CO_1\CO_3}=C_{\CO_3\CO_2\CO_1}$. Thus we have
\begin{align}
C_{\CO_1\CO_2\CO_3}=C_{\CO^\dagger_3\CO^\dagger_2\CO^\dagger_1}
\end{align}
On the other hand, parity invariance implies that 
\begin{align}
C_{\CO_1\CO_2\CO_3}=(-1)^{p_1+p_2+p_3}C_{\CO_2\CO_1\CO_3}
\end{align}
where $p_i$ is the parity of the field $\CO_i$.

We can easily show that the free Schr\"odinger field theory without any anti-particle (discussed later in Sec.~\ref{app4}) does not satisfy the parity invariance. We recall that the two point correlator on the $(\t,\vec{0})$ slice is given by
\begin{align}\label{freetwo}
\langle \phi(0)\phi^{\dagger}(\t)\rangle =\Theta(\t)\t^{-d/2}
\end{align}

If one assumes parity invariance with $\phi$ and $\phi^\dagger$ carrying opposite parity, the two point correlator satisfies:
\begin{align}
\langle \phi(0)\phi^{\dagger}(\t)\rangle = \langle \phi^\dagger(-\t)\phi(0)\rangle = \langle \phi(0) \phi^\dagger(-\t)\rangle\,.
\end{align}
Here the last equality follows because the cyclic order of the operator insertion is unaffected by \sl invariance. Thus $\langle \phi(0)\phi^{\dagger}(\t)\rangle$ has to be an even function of $\t$, which is not the case in \eqref{freetwo}. The presence of $\Theta(\t)$ implies the absence of anti-particles. 

On the other hand, when the parity symmetry is present one can write down a bootstrap equation \eqref{cross} even for a charged sector. The section~$4$ of \cite{Gaiotto:2013nva} elucidates on the scenario where a notion of parity is available. This motivates us to ask whether one can impose parity invariance on top of Schr\"odinger invariance. It indeed can be done by defining a free field theory such that the two point correlator takes the following form on $(\t,\vec{0})$ slice: 
\begin{align}
\langle \psi(0,\vec{0})\psi^{\dagger}(\t,\vec{0})\rangle = \frac{1}{|\t|^{d/2}}
\end{align}
where $N<0$ is the $U(1)$ charge carried by the field $\psi$. All the higher point correlators are determined by Wick contraction. On $(\t,0)$ slice, this theory is expected to behave like generalized Bose/Fermi theory in $0+1$ dimension \cite{Qiao:2017xif}. We remark that even if it is possible to impose parity invariance on $(\t,\vec{0})$ slice, it is not clear whether one can impose such invariance even away from the above mentioned temporal slice. This is because the boost invariance forces the $\vec{x}$ dependence of two point correlator to be $\exp\left(\imath N\tfrac{|\vec{x}|^2}{2t}\right)$ which is clearly not symmetric under $t\to-t$ unless one also imposes $N\to-N$ constraint. 
It would be interesting to consider another scenario, where one can have the following two point correlator:
\begin{align}
\langle \psi(0,\vec{0})\psi^{\dagger}(\t,\vec{x})\rangle = \frac{1}{|\t|^{d/2}}\exp\left[\frac{N|\vec{x}|^2}{2|\t|}\right]\,,
\end{align}
and it is obtained by different analytical continuation to imaginary time $\t$, depending on signature of real time $t$. It is not clear at present whether it carries any physical significance and leads to a well defined theory. Nonetheless, one can indeed have theories which enjoy parity invariance on the $(\t,\vec{0})$ slice.

\section{Free Schr\"odinger Field theory \& Its Euclidean avatar}\label{app4}

The free Schr\"odinger field theory in $d+1$ dimensions is described by
\begin{align}\label{action}
\mathcal{L}=2\imath \phi^{\dagger}\partial_t\phi + \phi^{\dagger}\nabla^2\phi\ ,
\end{align}
where $t$ is the real time and we have taken the mass to be unity. The propagator in momentum space representation has a pole at
\begin{align}
\omega =\frac{|\vec{k}|^2}{2}.
\end{align}
where $\omega$ is the energy and $\vec{k}$ is the momentum. This is in fact the dispersion relation of an on-shell particle, described the field theory~\eqref{action}. The presence of a pole brings in ambiguity in defining the position space propagator and thus necessitates a pole prescription. Similar ambiguity is present in the relativistic theory as well at free fixed point, where there are two poles at $\omega=\pm|\vec{k}|$. To circumnavigate the pole(s), one usually works in the imaginary time $\tau=\imath t, \omega_E=-\imath\omega$ (so called \textit{Wick rotation}), where the propagator is uniquely defined. Upon analytic continuation back to the Minkowski space-time, this provides us with a pole prescription. In what follows, we will follow the same procedure for the free Schr\"odinger field theory and come up with an expression for the propagator consistent with the \sl algebra.

The Wick rotated Schr\"odinger theory has a propagator of the following form
\begin{align}
\frac{1}{|\vec{k}|^2-2\imath\omega_E}
\end{align}
This does not have a pole on the real axis of $\omega_E$. Hence, the Fourier transform is well defined and unique and one obtains the Euclidean propagator $G_E$:
\begin{align}\label{prop}
\nonumber &G_{E}\left(\tau_1,\vec{x}_1;\tau_2,\vec{x}_2\right)=\int \frac{d\omega_E}{2\pi}\ \int \frac{d^dk}{(2\pi)^d}\ \frac{e^{-\imath(\omega_E(\tau_1-\tau_2)-\vec{k}\cdot\vec{x})}}{|\vec{k}|^2-2\imath\omega_E}\\
&=\frac{\Theta(\tau)}{2}\left(\frac{1}{2\pi\tau}\right)^{\frac{d}{2}}\exp\left(-\frac{\vec{x}^2}{2\tau}\right)
\end{align}
where $\tau=\tau_2-\tau_1$ and $\vec{x}=\vec{x}_2-\vec{x}_1$. Upon Wick rotation back to the real time, we have the following $\imath\epsilon$ prescription: 
\begin{align}\label{polep}
\frac{\imath}{2\omega-|\vec{k}|^2+\imath\epsilon}
\end{align}

\begin{figure}[!ht]
\includegraphics[scale=0.3]{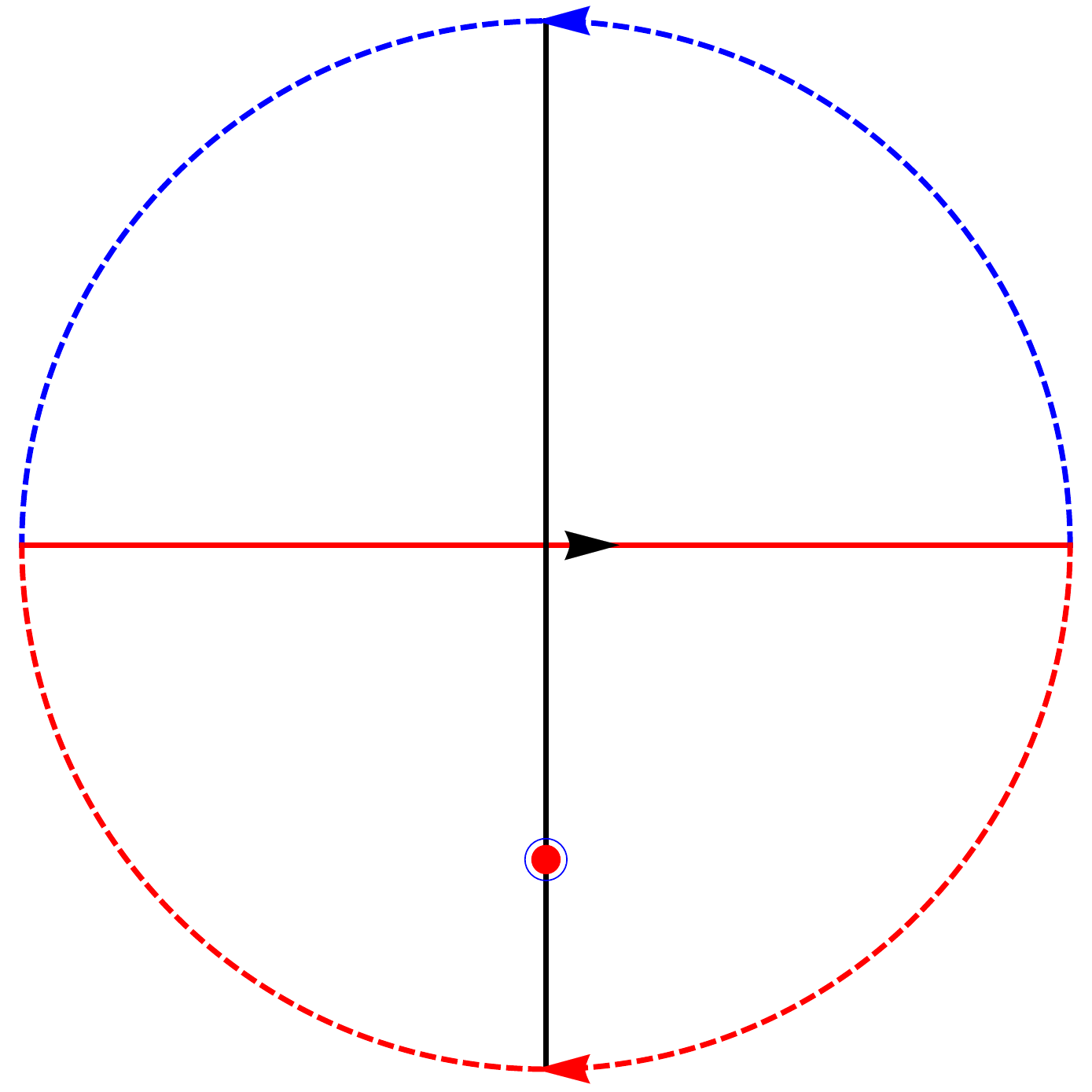}
\includegraphics[scale=0.3]{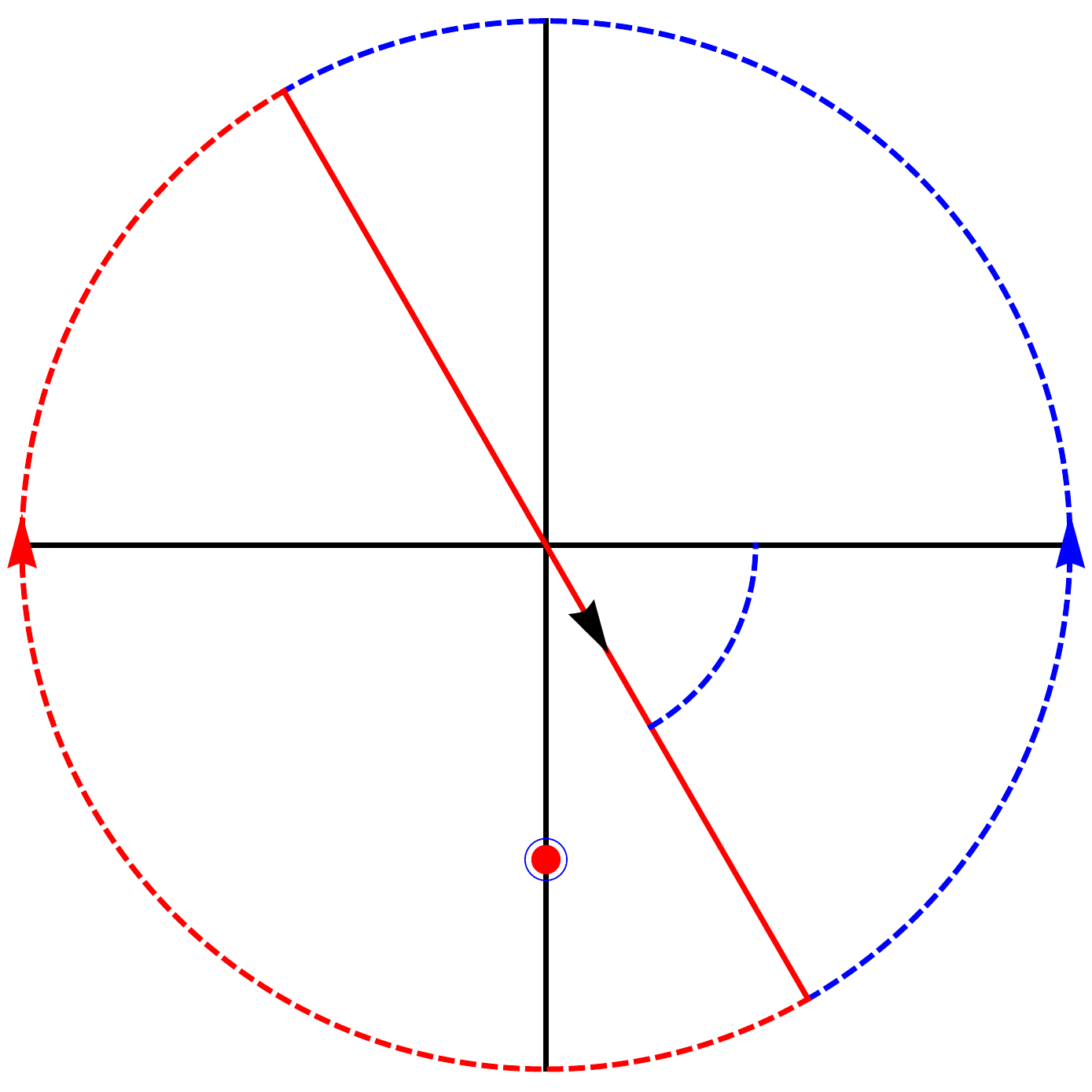}
\caption{Contour in the Wick rotated plane ($\omega_E$ plane)}
\label{Euclid2}
\end{figure}

\begin{figure}[!ht]
\includegraphics[scale=0.3]{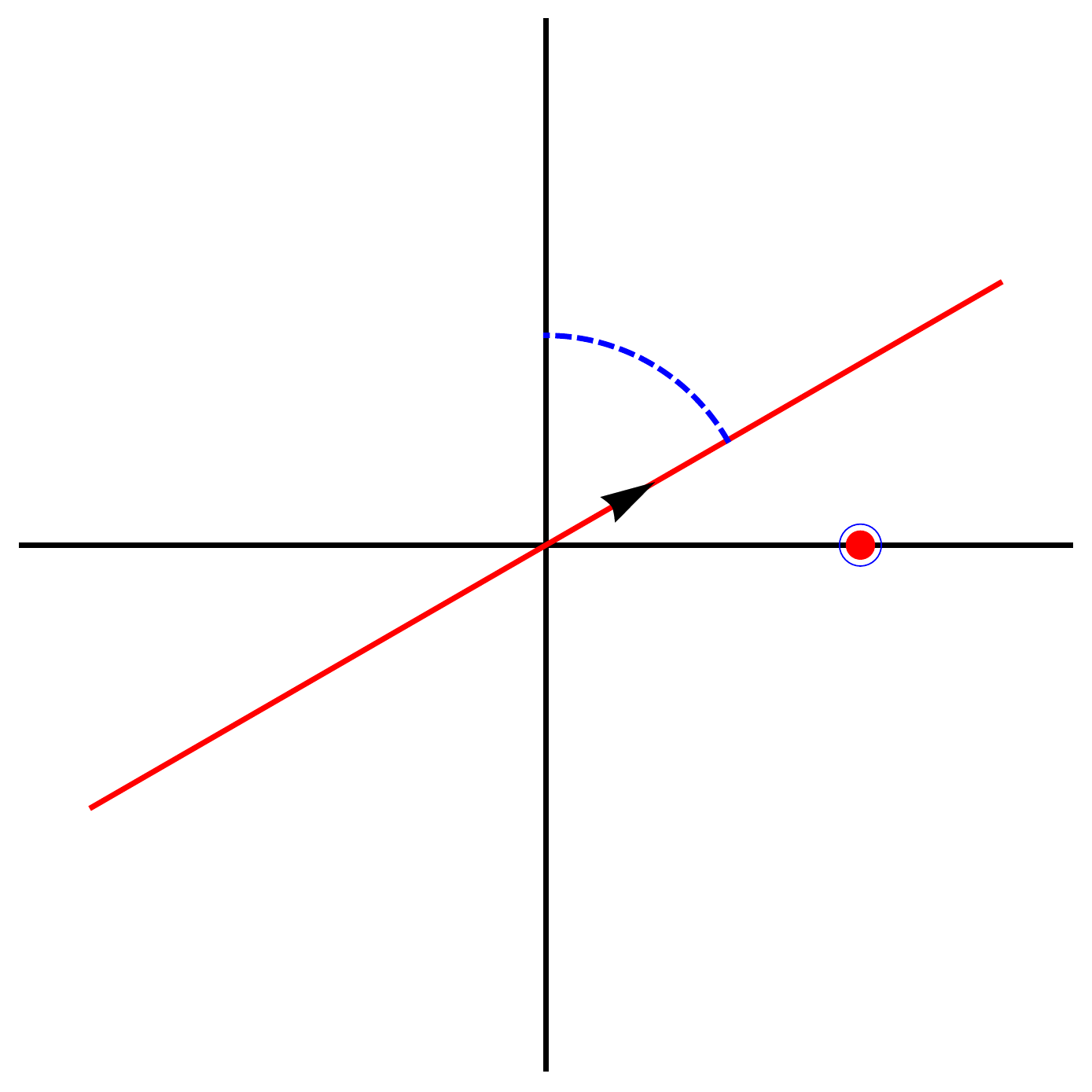}
\includegraphics[scale=0.3]{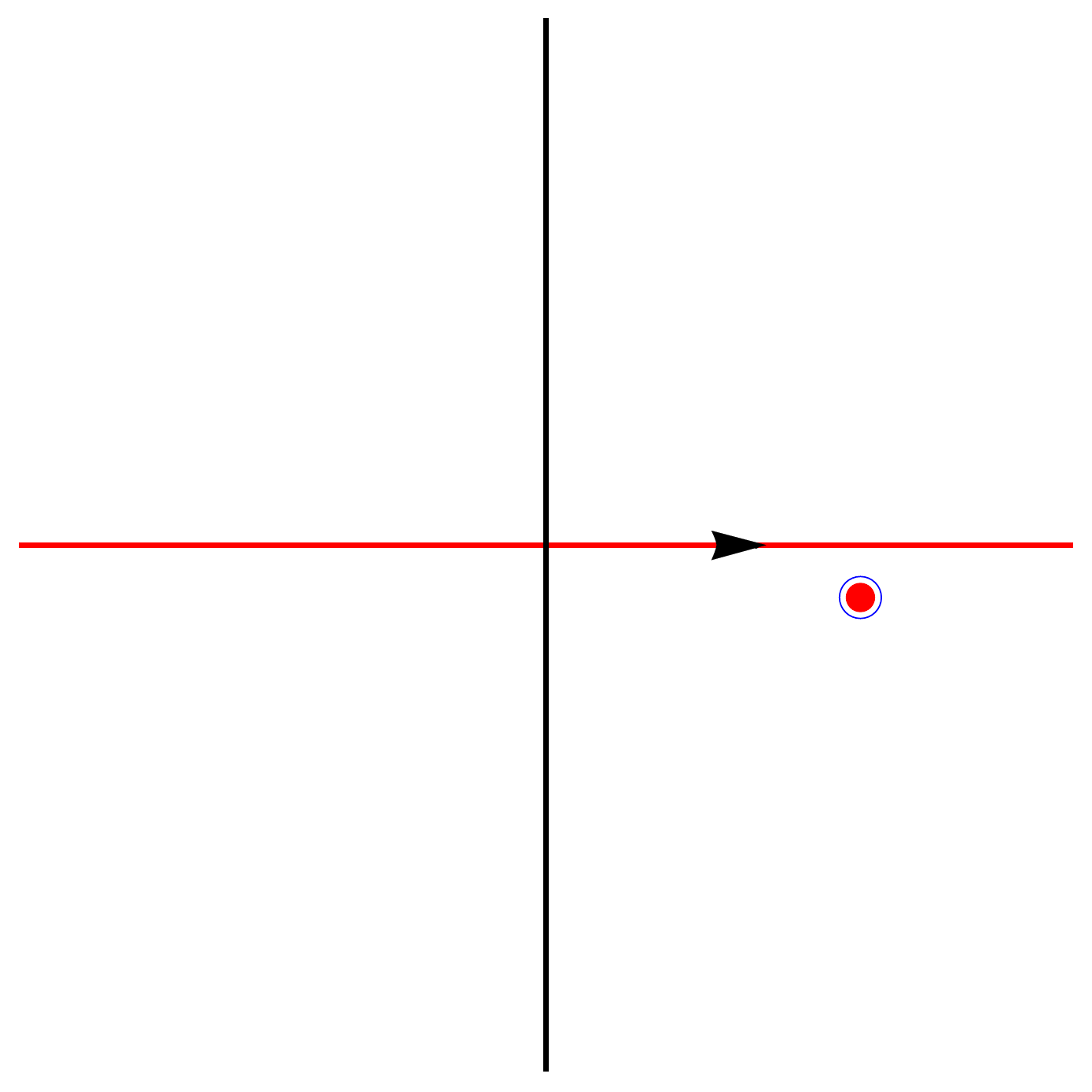}
\caption{Contour in the real $\omega$ plane}
\label{Mink2}
\end{figure}

The pole prescription, in momentum space, can be visualized through a series of diagrams, e.g. fig.~(\ref{Euclid2},\ref{Mink2}). We recall that the Wick rotation involves defining $\tau=\imath t $ and $\omega_E=-\imath\omega$ such that $e^{\imath\omega t}= e^{\imath\omega_E\tau}$ holds true. One can rotate the contour clockwise by an angle of $\pi/2-\epsilon$  on the $\omega_E$ plane, where $\epsilon$ is very small but a positive number, without affecting the integral (See fig.~\ref{Euclid2}). At this point, one effects the wick rotation, which recasts the integral and the contour as shown in the fig.~(\ref{Mink2}) and leads to the pole prescription, as in eq.~\eqref{polep}. 

Several remarks are in order. First of all, the physical significance of $\Theta(\tau)$ and hence, the $\imath\epsilon$ prescription lies in the fact that there are no antiparticles in the theory. This has consequences e.g. the vacuum does not have any spatial entanglement entropy, the weyl anomaly is absent upon coupling the theory with a non trivial Newton-Cartan structure\cite{Pal:2017ntk}. Furthermore, one can analytically continue the theory to live on a non integer dimensional space. The propagator as in eq.~\eqref{prop} can be defined by an analytic continuation in $d$. The analytical continuation is always understood to be in the number of spatial dimensions, without affecting the time co-ordinate.

The Schr\"odinger algebra constrains the real time two point correlator ($G$) of two primary operators of dimension $\Delta_{\mathcal{O}}$. It is given by 
\begin{align}\label{propM}
G(t_1,x_1;t_2,x_2)=ct^{-\Delta_{\mathcal{O}}}e^{\imath \frac{N_{\mathcal{O}}}{2} \frac{|\vec{x}|^2}{t}}
\end{align}
where $t=t_2-t_1$ and $\vec{x}=\vec{x}_2-\vec{x}_1$. We wick rotate the expression, introducing $\tau=\imath t$ and we choose $c=\frac{\Theta(\tau)}{2}\left(\frac{1}{2\pi\imath}\right)^{d/2}$ where $\tau=\tau_2-\tau_1$. Upon comparing this expression with the Euclidean propagator, as in eq.~\eqref{prop}, the dimension of free Schr\"odinger field is found out to be
\begin{align}
\Delta_{\phi}=\frac{d}{2}
\end{align}
which is precisely the engineering dimension as evident from the Lagrangian~\eqref{action}. 

We emphasize that the real time propagator, given in eq.~\eqref{propM}, is generic in the sense it is suitable to describe any Schr\"odinger invariant fixed point including the interacting ones. Thus the take home message is that the usual trick of Wick rotation is consistent, as it should be, with the notion of the Euclidean Schr\"odinger group and the Euclidean Schr\"odinger operator is indeed given by $(\partial_\t-\nabla^2)$.

\paragraph{Time ordering:}
The time-ordered (or anti-time ordered) real time correlator is obtained from the Euclidean correlator by proper analytic continuation. The Eq.~\eqref{prop} implies that
\begin{align}
\langle\phi(0)\phi^\dagger(\t)\rangle_{E}\propto \Theta(\t) 
\end{align}
where we have put in the subscript $E$ to clearly specify that it is a Euclidean correlator. Now we will show that the time ordered correlator can be obtained by taking $\tau=-\imath t$ while the anti-time ordered one can be obtained by taking $\tau=\imath t$. For $t>0$, we obtain the time ordered correlator by analytic continuation ($\tau=-\imath t$)
\begin{align}\label{c11}
 \langle 0|\phi^\dagger(t)\phi(0)|0\rangle\propto\Theta(-\imath t)=\imath \Theta(-t)=0\,, 
\end{align}
while for $t<0$, we obtain
\begin{align}\label{c21}
 \langle 0|\phi(0)\phi^\dagger(t)|0\rangle\propto\Theta(-\imath t)=\imath \Theta(-t)\neq 0\,.
\end{align}
Similarly, for $t<0$, the anti-time ordered correlator can be obtained by ($\tau=\imath t$)
\begin{align}\label{c31}
 \langle 0|\phi^\dagger(t)\phi(0)|0\rangle\propto\Theta(\imath t)=\imath \Theta(t)=0 
\end{align}
while for $t>0$, we have
\begin{align}\label{c41}
 \langle 0|\phi(0)\phi^\dagger(t)|0\rangle\propto\Theta(\imath t)=\imath \Theta(t)\neq 0 
\end{align}

It is easy to verify that all the equations~\eqref{c11},\eqref{c21},\eqref{c31},\eqref{c41} conforms to the fact that the field $\phi$ annihilates the vacuum, which is a manifestation of absence of anti-particles. We also remark that \eqref{c21} has interpretation of the amplitude associated with a particle being created at time $t<0$ and subsequently propagating to $t=0$. In the main text, we have used the analytic continuation $\t=\imath t$, one could have equivalently choose the other analytic continuation $\t=-\imath t$ and obtain similar results. \\

\paragraph{A different convention:}
The Euclidean time correlator can alternatively defined in following way where $\Theta(-\tau)$ appears in stead of \eqref{prop}:
\begin{align}\label{propalter}
\nonumber &G^{\text{alt}}_{E}\left(\tau_1,\vec{x}_1;\tau_2,\vec{x}_2\right)=\int \frac{d\omega_E}{2\pi}\ \int \frac{d^dk}{(2\pi)^d}\ \frac{e^{-\imath(\omega_E(\tau_2-\tau_1)-\vec{k}\cdot\vec{x})}}{|\vec{k}|^2-2\imath\omega_E}\\
&=\frac{\Theta(-\tau)}{2}\left(\frac{1}{2\pi\tau}\right)^{\frac{d}{2}}\exp\left(-\frac{\vec{x}^2}{2\tau}\right)
\end{align}
where $\tau=\tau_2-\tau_1$ and $\vec{x}=\vec{x}_2-\vec{x}_1$. With this convention, $\tau=\imath t$ will yield time-ordered correlator while $\tau=-\imath t$ will give the anti--time ordered one. As mentioned earlier, in the main text, we have adopted the convention as in \eqref{prop}.

\bibliography{refs}

\end{document}